# Contact Behaviour of Simulated Rough Spheres Generated with Spherical Harmonics

Deheng WEI[1], Chongpu ZHAI[2], Dorian HANAOR[3], Yixiang GAN[1]*

[1] School of Civil Engineering, The University of Sydney, Sydney, Australia

[2] Hopkins Extreme Materials Institute, John Hopkins University, Baltimore, United States

[3] Fachgebiet Keramische Werkstoffe, Technische Universität Berlin, Berlin, Germany

* Corresponding author: yixiang.gan@sydney.edu.au

**Highlights**

| | |
|---|---|
| i) | Fractal rough particles with multiscale topological features can be generated using an ultra-high degree Spherical Harmonic function. |
| ii) | Contact simulations show that contact stiffness presents a power law behaviour with the applied force, differing from Hertzian theory at low load. |
| iii) | Empirical relations between contact stiffness and load are proposed considering topological indices, including the relative roughness and fractal dimensions. |
| iv) | At the asperity scale, individual contact islands evolve and merge, following a Weibull-type distribution, independent of the loading level. |

**Abstract:** Normal contact behaviour between non-adhesive fractal rough particles was studied using a finite element method (FEM). A series of spherical grain surfaces with distinguished roughness features were generated by means of Spherical Harmonics. These surfaces were described by two roughness descriptors, namely, relative roughness ($R_r$) and fractal dimension ($FD$). The contact behaviour of rough spheres with a rigid flat surface was simulated using FEM to quantify the influences of surface structure and sphere morphology by focusing on contact stiffness and true contact area. The dependence of normal contact stiffness ($k$) on applied normal force ($F$) was found to follow a power law ($k = \alpha F^{\beta}$) over four orders of magnitude, with both $\alpha$ and $\beta$ being highly correlated with $R_r$ and $FD$. With increasing load, the power exponent converges to that of Hertzian contact, e.g., 1/3, independent of $R_r$. Regions of true contact evolved through the formation of new microcontacts and their progressive merging, meanwhile the area distributions of contact island induced by various forces tend to obey similar Weibull distributions due to fractal nature in their surfaces. Contacts with larger values of $R_r$ were found to produce contact contours with higher fractal dimension as calculated by a 2D box-counting method. Our results suggest that the correlation between radial lengths in a quasi-spherical particle should be considered in studying contact behaviour.

## 1. Introduction

Granular materials are ubiquitous on Earth and account for a significant portion of the landmass occurring on Earth's surface. The physical behaviour of granular systems may exhibit characteristics of solids, gas or liquids (Jaeger and Nagel, 1990). As implied by William Blake, who famously wrote 'to see a world in a grain of sand', nearly all of the laws of physics can be observed in granular matter. However, the observation of a specific phenomenon depends on measurement resolution, due to the geometrical and topological variation of granular materials across multiple scales. In the context of soil mechanics, particle morphology can be categorized into three diverse yet correlated length scales, namely (1) aspect ratio, roundness and roughness for particulate size, (2) local structure and angularity, and (3) asperity structures down to the finest scales (Wadell, 1932; Barrett, 1980). Across multiple scales, various correlations between morphology and system behaviour have been established using diverse experimental techniques, including stereophotography (Zheng and Hryciw, 2017; Sun et al., 2019), photoelastic (Dantu, 1957; Hurley et al., 2014), and X-ray computed tomography (Petrovic et al., 1982; Viggiani et al., 2004), and observing phenomena such as critical state (Roscoe et al., 1958; Yang and Luo, 2015), stress-dilatancy relation (Taylor, 1948; Li and Dafalias, 2000), and the evolution of fabric tensor (Oda, 1972; Gao and Zhao, 2013). However, as the foundation of granular mechanics, the contact behaviour of natural grains at the particle scale (Zhai et al., 2019) merits further investigation. Alternatively, some existing descriptions have been established within the context rigid-particle approximation (Khun and Bagi, 2004; Li et al., 2011; Kuhn and Daouadji, 2018) from Discrete Element Method (DEM).

DEM was firstly established in the realm of soil mechanics, as an approach in which rigid volume-equivalent spheres are taken as first-order approximations for irregular shaped sand particles in order to facilitate computationally efficient simulations of granular behaviour (Cundall and Strack, 1979). Over the past four decades, DEM has been significantly developed by considering more complicated particle shapes, for example, clusters of discs or spheres (de Bono and McDowell, 2018), ellipsoids (Ng et al., 2018), super or poly super ellipsoids (Zhao et al., 2018; Zhao and Zhao, 2019), polyhedrons (Latham and Munjiza, 2004) and also realistic shapes based on image processing and mathematical model (Andrade et al., 2012; Mollon and Zhao, 2014; Kawamoto et al., 2016). Although complicated shapes can be used in DEM, the calculation of contact forces is the product of progressively overlapping lengths and instant contact stiffness provided by the input contact law, thus, only qualitative analysis of particulate systems can be achieved. Discrepancy will appear between DEM simulation and real observations (Cavarretta et al., 2010; Zhai et al., 2016a; Nardelli and Coop, 2018) especially for small-strain related geotechnical problems including wave propagation and liquefaction (Yimsiri and Soga, 2000; Chang and Hicher, 2005). These inconsistences may be traced back to the elastic contact region where fine-detailed surface features dominate contact stiffness.

Besides contact stiffness (Persson, 2006; Akrapu et al., 2011; Pohrt and Popov, 2012), real contact area plays a determining role in many other physical phenomena, such as electrical conduction (Yastrebov et al., 2015; Zhai et al., 2015, 2016b) and thermal transport (Owen and Thomson, 1963; Brutsaert, 1975; Persson et al., 2010). While numerous studies can be found on contact between nominally flat rough surfaces, the contact behaviour of curved rough surfaces (e.g., granular materials in DEM) has attracted considerably less attention. As early as 1882, Hertz analytically studied frictionless and nonadhesive contact between a rigid smoothed sphere or asperity and an elastic flat surface (Johnson, 1985). Greenwood and Tripp (GT model, 1967) first modelled the elastic contact between rough spheres based on the prevalent Greenwood-Williamson (GW) model (Greenwood and Williamson, 1966), where contact between two rough surfaces can be seen as the collection of asperity contacts with a nominally flat surface. Due to inherent limitations of the original GW model discussed elsewhere (Barber and Ciavarella, 2000; Greenwood and Wu 2001; Vakis et al., 2018), GT model uses only standard deviations of sphere 'radii' distributions to quantify roughness globally without considering their correlation. For greater simplicity, the contact of two elastic rough surfaces can be considered equal to that between an equivalent rough surface and a flat (Johnson, 1985; Barber, 2003), which has been applied by nearly all simulations of contacting rough surfaces. As for rough spheres, usually roughness is only mapped into the sphere or corresponding flat platen, and only one of them is elastic and the other being rigid (Kagami et al., 1983; Cohen et al., 2009; Pohrt and Popov, 2013; Pastewka and Robbins, 2016; Yastrebov, 2019), insufficiently covering the influences of roughness asperities if roughness and elasticity are not present on the sphere at the same time.

Many experimental approaches have been developed for characterizing surface morphology, including nano-CT (Shearing et al., 2010), near-field diffraction (Nomura et al., 2005), laser profilometry (Weber et al., 2018), atomic force microscopy (AFM) (Buzio et al., 2003) interferometry (Ovcharenko et al., 2006), frustrated total internal reflection (Rubinstein et al., 2004), and phase-contrast microscopy (Dyson and Hirst, 1954). However, in these methods the global 3D object morphology is limited by lateral or vertical resolution, requiring the simplified simulation of finest scale features in many computationally generated rough surfaces, by methods such as power spectrum density (Persson et al., 2002; Yastrebov et al., 2015; Müser, 2018) and Weierstrass-Mandelbrot (Chiaia, 2002; Ciavarella et al., 2006; Hanaor et al., 2015) functions. To improve the estimation of the standard deviation of radial length distributions for rough spheres in the aforementioned GT model, the power spectrum is directly mapped onto a sphere for numerical analysis of rough sphere contact behaviours (Pohrt and Popov, 2013; Pastewka and Robbins, 2016), however, with the increase of vertical height, the mesh size becomes large, gradually becoming unrealistic in computational frameworks for contact mechanics, such as Boundary Element Method (BEM) and FEM. Hence, it is reasonable to separate fine surface features from the global particle morphology, through two steps: first the reconstruction of particle morphology, on the basis of experimental data at the global scale, and then the inclusion of a constant fractal dimension (*FD*) at finer scales. The key of the latter step is to implement an efficient method to quantify particle *FD* using limited experimental data at the surface scale. As early in 1977, Meloy implemented Fast Fourier Transformation (FFT) to reconstruct 2D particle outline and found the logarithmic linear relation between Fourier descriptors and degrees (Meloy, 1977), which was further proved by Bowman et al. (2000) and Mollon and Zhao (2012). Recently, the 2D outline has been extended to 3D surface via 3D FT, Spherical Harmonic function (Wei et al., 2018), with the aid of X-ray computed tomography (CT).

In the present work we revisit the contact behaviour of rough spheres using a more versatile FEM approach, motivated by the identified shortcomings of existing analytical approaches, which generally consider spherical grains while neglecting asperity deformation.   (Hyun et al., 2004; Etsion et al., 2005; Pei et al., 2005; Yastrebov, 2013; Xu et al., 2015). This study is organized as follows. Section 2 provides a brief recapitulation of Spherical Harmonics to reconstruct or smooth particle shapes, and a thorough application of extremely high degrees for multi-scaled sphere morphology for providing fine-scale feature on a curved surface. In addition, the necessary details of FEM simulation, for particle contact behaviours using graded mesh, are provided. Then radial length and mean curvature distributions of the generated rough sphere are discussed. Section 3 describes the results of contact mechanics simulations, with the focus on the combined effects from surface curvature and roughness, i.e., global and local features. Contact stiffness, contact area, and radial contact stress distributions are described. Discussions and conclusions are presented in Section 4.

## 2. Methodology

This section includes two main parts, namely generating spheres with rough surfaces from a smooth sphere and finite element method (FEM). The rough spheres can be generated using Ultra-high-degree Spherical Harmonics technique (e.g. SH degree up to 2000) to reproduce morphological surface features coexisting with a global curvature. The outline of the FEM modelling of rough surface contact is illustrated in Fig. 1. For FEM simulations, the result is first validated by a Hertzian contact solution mimicking the contact behaviour between an elastic sphere, with graded mesh sizes, and a rigid flat plane. Then, later in Section 3, the contact behaviour of rough spheres is investigated focusing on the dependency on the morphological variations.

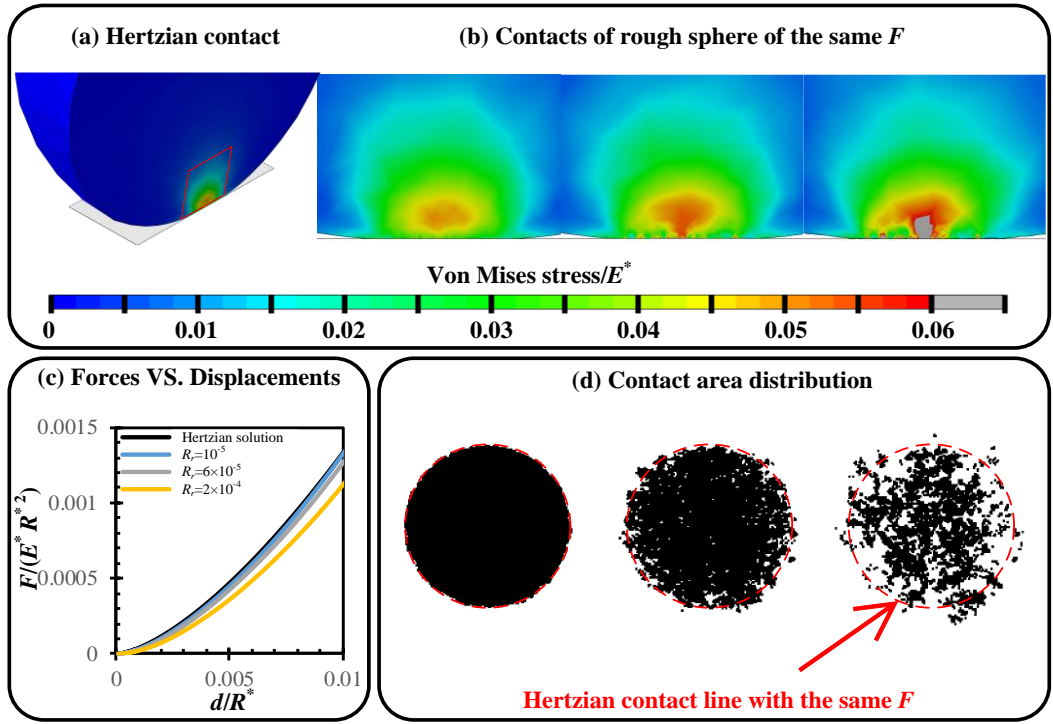

**Fig. 1** Outlines of FEM simulations of rough sphere contact behaviour.

## 2.1 Ultra-high Spherical Harmonics (SH) for morphology features

This Section demonstrates the efficiency of SH-based fractal dimension (*FD*) in characterizing particle morphology. SH descriptors have been previously used for describing particle shapes (e.g., Wei et al., 2018) and the potential of extending this to fine-scale roughness is further explored here.

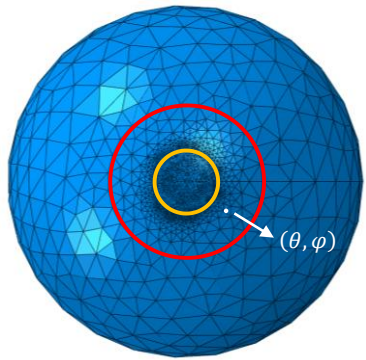

**Fig. 2** Graded FEM mesh for the smooth sphere in the spherical coordinate system.

A surface point on a star-like particle can be represented in terms of its distance from the particle centroid $r_i(x_i(\theta, \varphi), y_i(\theta, \varphi), z_i(\theta, \varphi))$ in a polar coordinate system, as in Fig. 2, by the orthogonal SH function:

$$r_i(\theta, \varphi) = \sum_{n=0}^{\infty} \sum_{m=-n}^{n} c_n^m Y_n^m(\theta, \varphi), \tag{1}$$

$$r_i(\theta,\varphi) = \sqrt{(x_i - x_0)^2 + (y_i - y_0)^2 + (z_i - z_0)^2}, \tag{2}$$

where $i$ corresponds to $i$-th selected points on the particle surface, $(x_i, y_i, z_i)$ and $(x_0, y_0, z_0)$ are the Cartesian coordinates of the surface point and the chosen center inside the particle, $\theta \in [0,\pi]$ and $\varphi \in [0,2\pi)$ are the latitudinal and longitudinal coordinates respectively, and $c_n^m$ are the SH coefficients to be determined of degree $n$ and order $m$; Similar to Fourier series to represent functions on a planar circle, $Y_n^m(\theta,\varphi)$ ($n \in N$, $-n \leq m \leq n$) is the so-called SH function, the angular solutions of Laplace's equation for organizing spatial angular frequency, and defined on the surface of a sphere as:

$$Y_n^m(\theta,\varphi) = \sqrt{\frac{(2n+1)(n-|m|)!}{4\pi(n+|m|)!}} P_n^m(\cos\theta) e^{im\varphi}, \tag{3}$$

$$Y_n^{-m}(\theta,\varphi) = (-1)^m \left[\sqrt{\frac{(2n+1)(n-|m|)!}{4\pi(n+|m|)!}} P_n^m(\cos\theta) e^{im\varphi}\right]^*, \tag{4}$$

where $[.]^*$ denotes the complex conjugate and $P_n^m(x)$ are called associated Legendre functions, which can be expressed by Rodrigues' formula:

$$P_n^m(x) = (1-x^2)^{\frac{|m|}{2}} \cdot \frac{d^{|m|}}{dx^{|m|}}\left[\frac{1}{2^n n!} \cdot \frac{d^n}{dx^n}(x^2-1)^n\right]. \tag{5}$$

For the single degree $n$, there are $2n+1$ complex numbers of SH coefficients to be determined according to Eq. (1), hence when the user-defined maximum degree is $n_{max}$, the whole set of SH coefficients $c_n^m$ includes $(n_{max}+1)^2$ complex numbers for representing a 3D surface. Due to the orthonormal properties of SH function, via choosing the angles $(\theta, \varphi)$ at Gaussian quadratures the more general calculation of $c_n^m$ to reconstruct or smooth target particle shapes (Garboczi and Bullard, 2017) follows the integral

$$c_n^m = \int_0^{2\pi} \int_0^{\pi} \sin\theta \cdot r(\theta,\varphi) \cdot [Y_n^m(\theta,\varphi)]^* d\theta d\varphi. \tag{6}$$

In general, a greater degree of SH expansion corresponds to the representation of finer features of particle morphology.

The amplitude at each SH frequency can be measured by:

$$L_n = \sqrt{\sum_{m=-n}^{n} \|c_n^m\|^2}, (n = 0 \cdots 15), \tag{7}$$

where $\|.\|$ is the second-order norm. To further quantify the development rule of the amplitudes at different SH frequencies, $L_n$ values were normalized by $L_0$ to eliminate the influence of particle volume. Moreover, because $L_1$ does not influence the SH-reconstructed particle morphology, $L_1$ was not considered. The SH descriptors characterizing the particle morphology can be finally defined as:

$$\begin{cases} D_0 = 1 \\ D_n = L_n/L_0, (n = 2, 3, 4, 5 \ldots) \end{cases} \tag{8}$$

The exponential relation between SH descriptor $D_n$ and SH degree $n$ can be expressed by:

$$D_n \propto n^\beta, \quad (9)$$

where $\beta = -2H$ is the slope of the regression plot of log ($D_n$) versus log ($n$) and $H$ is the Hurst coefficient that is related to the Fractal Dimension (*FD*) of Fourier transformation (Quevedo et al., 2008; Russ, 2013) by the following expression:

$$FD = 3 - H = (6+\beta)/2 \quad (10)$$

Appendix I details the logarithmic relations between SH descriptor $D_n$ and degree $n$ of six kinds of particles. Again, it is proven that SH-based *FD* enables the description of hierarchical particle morphological features.

To calculate the difference between two objects, according to Parseval's theorem and orthogonality of SH function,

$$\int_0^{2\pi} \int_0^{\pi} r(\theta, \varphi)\, d\theta d\varphi = \frac{1}{4\pi} \sum_{n=0}^{\infty} \sum_{m=-n}^{n} \|c_n^m\|^2 \quad (11)$$

Root Mean Square Distances (*RMSD*), associated with their SH coefficients, between two objects can be applied to quantify how globally different they are and follows (Gerig et al., 2001; Shen et al., 2009):

$$RMSD = \sqrt{\frac{1}{4\pi} \sum_{n=0}^{n_{max}} \sum_{m=-n}^{n} \|c_{1,n}^m - c_{2,n}^m\|^2}, \quad (12)$$

where $c_{1,n}$ and $c_{2,n}$ are the SH coefficients of two surfaces. Due to the characteristic length scale (i.e. asperity) of contacts of rough particles, extremely high degree of SH is necessary. Due to the cumulative changes of each order $m$, the closed form bridging wavelength, in terms of angular resolution, and the specific degree $n$ cannot be specified directly. Instead, a frequently-used rule of thumb is given by (Jekeli, 1996):

$$\Delta\theta = \frac{\pi}{n}, \quad (13)$$

where $\Delta\theta$ is the polar angle between two adjacent surface points. Furthermore, Fig. 3 shows a rough sphere's surface depicted using different degrees of SH expansion and the same minimum SH degree from identical sets of SH coefficients. It is evident that for simulating asperities of rough spheres the maximum SH degree should be higher. From analogous FEM simulations of contact between rough flats of unit square surface (Hyun et al., 2004; Pei et al., 2005), a fine mesh size 1/128 of the global surface square is found to be sufficient. In this study, the normalized maximum separation or overlap by sphere radius is roughly equal to 0.02, hence the selected mesh ratio (0.0015) is considered to be fine enough, as shown in the zone of refined FE meshes in Fig. 2, and the further proof will be discussed hereafter. Consequently, according to Eq. (13) and Fig. 3 the maximum SH degree is set to 2000, and $c_{2,n}$ is set to the SH coefficient of an unit sphere:

$$c_{2,n} = \begin{pmatrix} c_{2,0} \\ c_{2,1} \\ c_{2,2} \\ \vdots \\ c_{2,n_{max}} \end{pmatrix}^T = \begin{pmatrix} 2\sqrt{\pi} \\ \mathbf{0} \\ \mathbf{0} \\ \vdots \\ \mathbf{0} \end{pmatrix}^T = \begin{pmatrix} 2\sqrt{\pi} \\ (0 \quad 0 \quad 0)^T \\ (0 \quad 0 \quad 0 \quad 0 \quad 0)^T \\ \vdots \\ \underbrace{(0 \quad \cdots \quad 0)^T}_{2\times n+1} \end{pmatrix}^T. \tag{14}$$

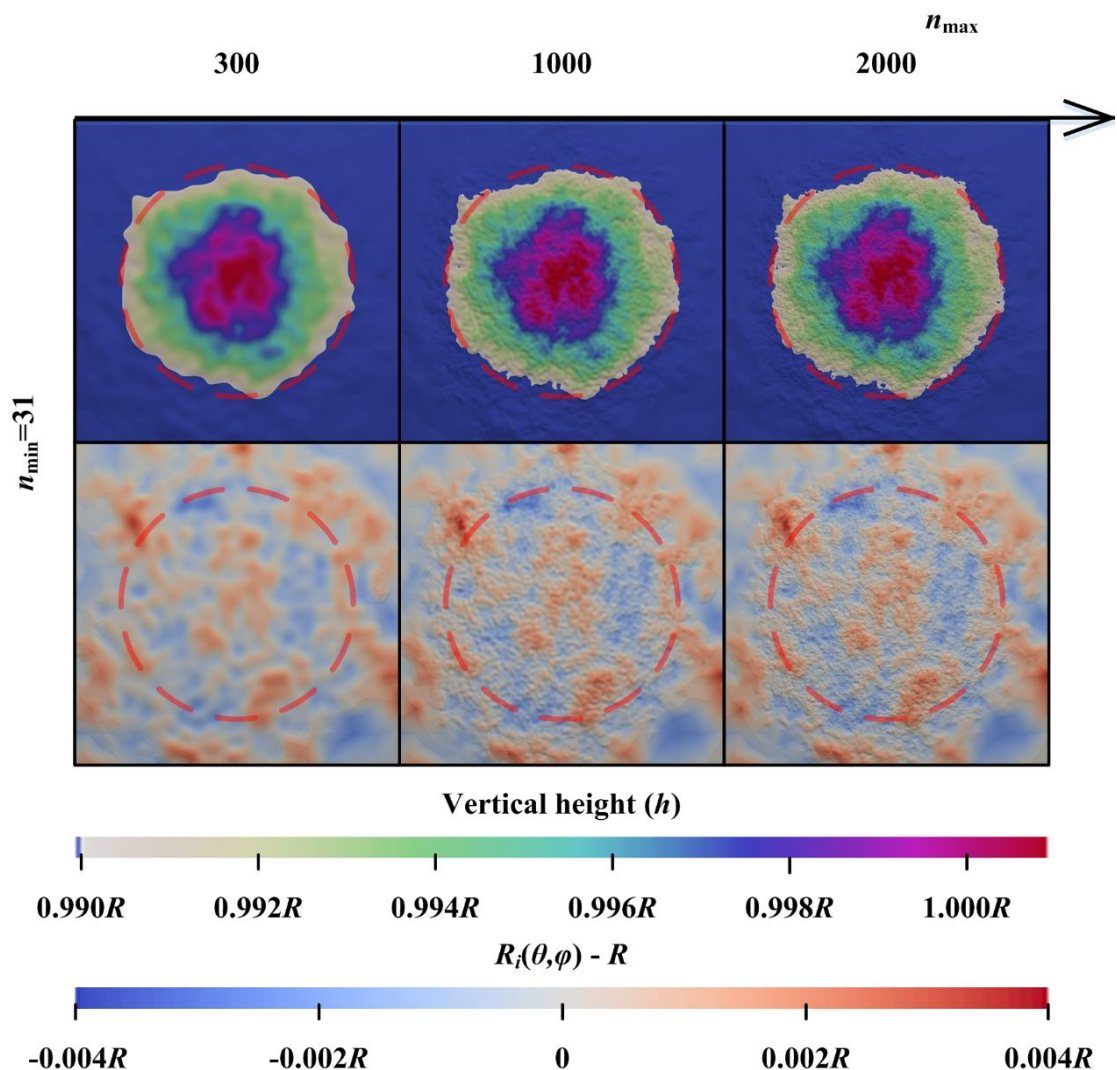

**Fig. 3** Rough sphere morphology features determined by different degrees of the same set of SH coefficients (the red dashed line denotes the boundary of the mesh refinement zone in FEM).

By conducting SH expansion to a greater degree, finer and finer details of particle surfaces can be depicted. To isolate roughness from roundness or curvature in the study of contact behaviour, rough spheres are considered with isotropic macro-scale curvature. It is possible to combine global (e.g., roundness and curvature) and local (e.g., roughness) features using a complete set of SH descriptors. However, this work focuses on the roughness features of a curved surface, to highlight the competing effects of local roughness and curvature. Hence, the lower SH degree was here set to 31, which is high enough to serve as the cut-off between roundness and roughness (Garboczi, 2002; Zhao et al., 2017). Considering $31 \leq n \leq 2000$ and 31, the ratio of large to small wavelengths are about 65, which is of the same magnitude of that (512/4) of molecular simulations in Pastewka and Robbins (2016). Meanwhile, we apply the power spectrum density, widely used in nominally flat surfaces, to directly quantify the cut square area with protection of size $0.2R \times 0.2R$ at mesh fine zones of SH-generated rough sphere surfaces. The log-log linear segment, ranging over more than two orders of magnitude in the calculated power spectrum densities, demonstrates self-affine features of the SH-generated rough structures. After applying roll-off and cut-off of wavelengths in terms of the SH degree, SH coefficients $c_{1,n}$ can be explicitly denoted:

$$c_{1,n} = \begin{pmatrix} c_{1,0} \\ c_{1,1} \\ c_{1,2} \\ \vdots \\ c_{1,30} \\ c_{1,31} \\ \vdots \\ c_{1,2000} \end{pmatrix}^T = \begin{pmatrix} 2\sqrt{\pi} \\ (0 \quad 0 \quad 0)^T \\ (0 \quad 0 \quad 0 \quad 0 \quad 0)^T \\ \vdots \\ \underbrace{(0 \quad \cdots \quad 0)^T}_{2\times 30+1} \\ \underbrace{\left(c_{1,31}^{-31} \quad \cdots \quad c_{1,31}^{0} \quad \cdots \quad c_{1,31}^{31}\right)^T}_{2\times 31+1} \\ \vdots \\ \underbrace{\left(c_{1,2000}^{-2000} \quad \cdots \quad c_{1,2000}^{0} \quad \cdots \quad c_{1,2000}^{2000}\right)^T}_{2\times 2000+1} \end{pmatrix}^T . \quad (15)$$

The production of virtual complex SH coefficients in polar coordinate systems has been conducted in earlier studies (Wei et al., 2018). However, the capacity ($10^{-307}$, $10^{307}$) of standard 64-bit computers can be rapidly exceeded. i.e. when $n$ is higher than 200 according to Eqs. (3) and (4) in the calculation of SH functions. We applied recursion formulae of $\sqrt{\frac{(2n+1)(n-|m|)!}{4\pi(n+|m|)!}} P_n^m(\cos\theta)$, set to $\overline{P_n^m}(\cos\theta)$, to approximate the SH function (see Page 963 in Gradshteyn and Ryzhik, 2007):

$$\begin{cases} \overline{P_0^0}(\cos\theta) = 1 \\ \overline{P_1^1}(\cos\theta) = \sqrt{3}\sin\theta \\ \overline{P_n^m}(\cos\theta) = \alpha_n^m \cos\theta \cdot \overline{P_{n-1}^m}(\cos\theta) - \beta_n^m \overline{P_{n-2}^m}(\cos\theta), n \geq 2, 0 \leq m \leq n-2 \\ \overline{P_n^{n-1}}(\cos\theta) = \sqrt{2n+1}\cos\theta \cdot \overline{P_{n-1}^{m-1}}(\cos\theta), n \geq 1 \\ \overline{P_n^n}(\cos\theta) = \sqrt{\frac{2n+1}{2n}}\sin\theta \cdot \overline{P_{n-1}^{n-1}}(\cos\theta), n \geq 2 \\ \alpha_n^m = \sqrt{\frac{(2n+1)(2n-1)}{(n+m)(n-m)}} \\ \beta_n^m = \sqrt{\frac{(2n+1)(n+m-1)(n-m-1)}{(2n-3)(n+m)(n-m)}} \end{cases}$$

(16)

Although a recursion formula is adopted here, a data underflow phenomenon can still appear in some regions. For example, when for low angular separations ($\theta = 0$ or $\pi$), The value of $\overline{P_{2500}^{2500}}(\cos\theta)$ is about $10^{-5000}$. Hence, a scaling factor $Q=10^{260}$ is introduced to $\overline{P_n^m}(\cos\theta)$, by which arbitrary polar angles can be computed up to the maximum SH degree $n = 2190$ higher than 2000 for reconstructing rough spheres in this study. Simultaneously, to remove influences of randomness of complex SH coefficients on particle morphology, all the rough sphere surfaces are generated by the same set of SH coefficients $c_n^m$ multiplying real numbers $k_n$ to form the relations between $n$ and $D_n$ to introduce fractal dimension in Fig. 4, hence Eq. (1) can be written as:

$$r_i(\theta,\varphi) = \frac{1}{Q}\sum_{n=0}^{\infty}\sum_{m=-n}^{n} k_n c_n^m [Q\overline{P_n^m}(\cos\theta)e^{im\varphi}], k_n \in \mathbb{R}, \qquad (17)$$

where $\mathbb{R}$ denotes real number. Furthermore, relative roughness ($R_r$) is defined as the normalized $RMSD$ by radius of unit sphere from $c_{2,0}$ in Eq. (14):

$$R_r = \frac{\sqrt{\frac{1}{4\pi}\sum_{n=0}^{n_{max}}\sum_{m=-n}^{n}\|c_{1,n}^m - c_{2,n}^m\|^2}}{c_{2,0}^0 \cdot Y_0^0(\theta,\varphi)}. \qquad (18)$$

In practice, this term describes the ratio between the local roughness and the radius of surface curvature. Then according to Eqs. (7) to (10) and (18), when $FD$ and $R_r$ are given, $k_n$ can be determined, followed by $D_n$ and $L_n$.

Notably, the selection of polar angles of refined areas in Fig. 2 can also vary the actual spatial distribution of asperities even for the same SH coefficients. Here, the same reference polar angle $(\theta,\varphi)$, with a nominally lowest point equal to $(\frac{3\pi}{2},\frac{\pi}{2})$ in Fig. 2, is applied for different rough spheres. Fig. 4 illustrates relations between $D_n$ and $n$ of generated rough spheres, while Fig. 5 represents the distributions of radius lengths and curvature values of $R_r$ equal to $10^{-5}$ and $2\times10^{-4}$. Rough surfaces are often characterised in terms of standard deviations of height distributions in rough surfaces (Persson et al., 2002) or radial length distributions in the case of rough spheres (Greenwood and Tripp, 1967), since these are mostly Gaussian distributions, this corresponds to commonly used terms of roughness. In this study, the normalized roughness of mean radial length For $R_r=10^{-5}$ and $2\times R_r=10^{-5}$ are about $2\times10^{-5}$ and $10^{-3}$, respectively. The normalized spherical roughness in analytical solutions of Greenwood and Tripp (1967) corresponds to [$2\times10^{-5}$ $2\times10^{-4}$]. Hence, it is reasonable to separate contact response into three stages according to normal contact force via GT model in the following parts. Although the radial length distributions of Fig. 5 (a) and (b) nearly coincide, their morphological features indicated by the contour maps are divergent, meanwhile (c) and (d) demonstrate that larger values of $R_r$ increase $FD$ and the mean curvature (see Appendix II) value distributions. Greenwood and Tripp (1967) applied a deterministic analytical solution (Greenwood and Williamson, 1966) to investigate elastic behaviours of rough spheres and concluded that the response, influenced by geometry, is governed by asperity density, roughness (the standard deviation of radius length distribution taken to be a strict normal distribution) and sphere-shaped asperity curvature. Interestingly, normal distributions can describe both radial length and mean curvature value distributions of SH-based fractal surfaces of spheres in Fig. 5. Regarding the approximation of asperity shape, Ciavarella et al. (2006) re-vitalized the Greenwood and Williamson model and considered the error of mean asperity curvature for Gaussian surfaces to be of a constant order. However, from the contour maps of Fig. 5 (c) and (d) the mean curvature may not be always of the same order, and the scope would be much larger when Gaussian curvature values are taken instead. Especially for explicit numerical simulations and experiments, the points between or connecting asperity regions may also make contact with the compressing platen, although they are not commonly considered as asperities. Our results here are based on logarithmic linear relations between $D_n$ and $n$ of SH analysis, which was experimentally confirmed in larger-length scale, but assumed in finer morphology details.

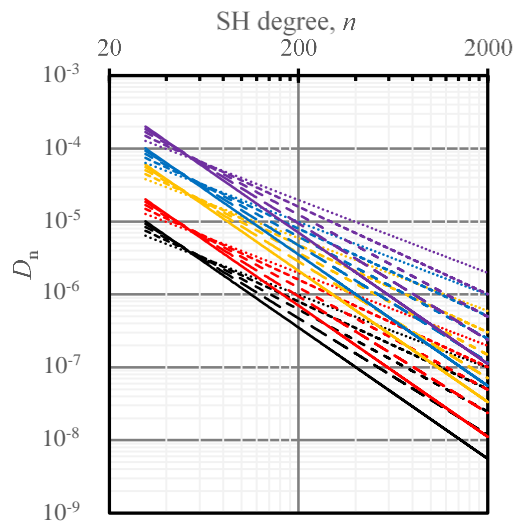

**Fig. 4** Relations between $D_n$ and SH degree of generated rough spheres. Colour black, red, yellow, blue and purple indicate $R_r$ equal to $10^{-5}$, $2\times10^{-5}$, $6\times10^{-5}$, $10^{-4}$ and $2\times10^{-4}$; the scattered levels of lines mean *FD* equal to 2.0, 2.1, 2.2, 2.3, 2.4 and 2.5.

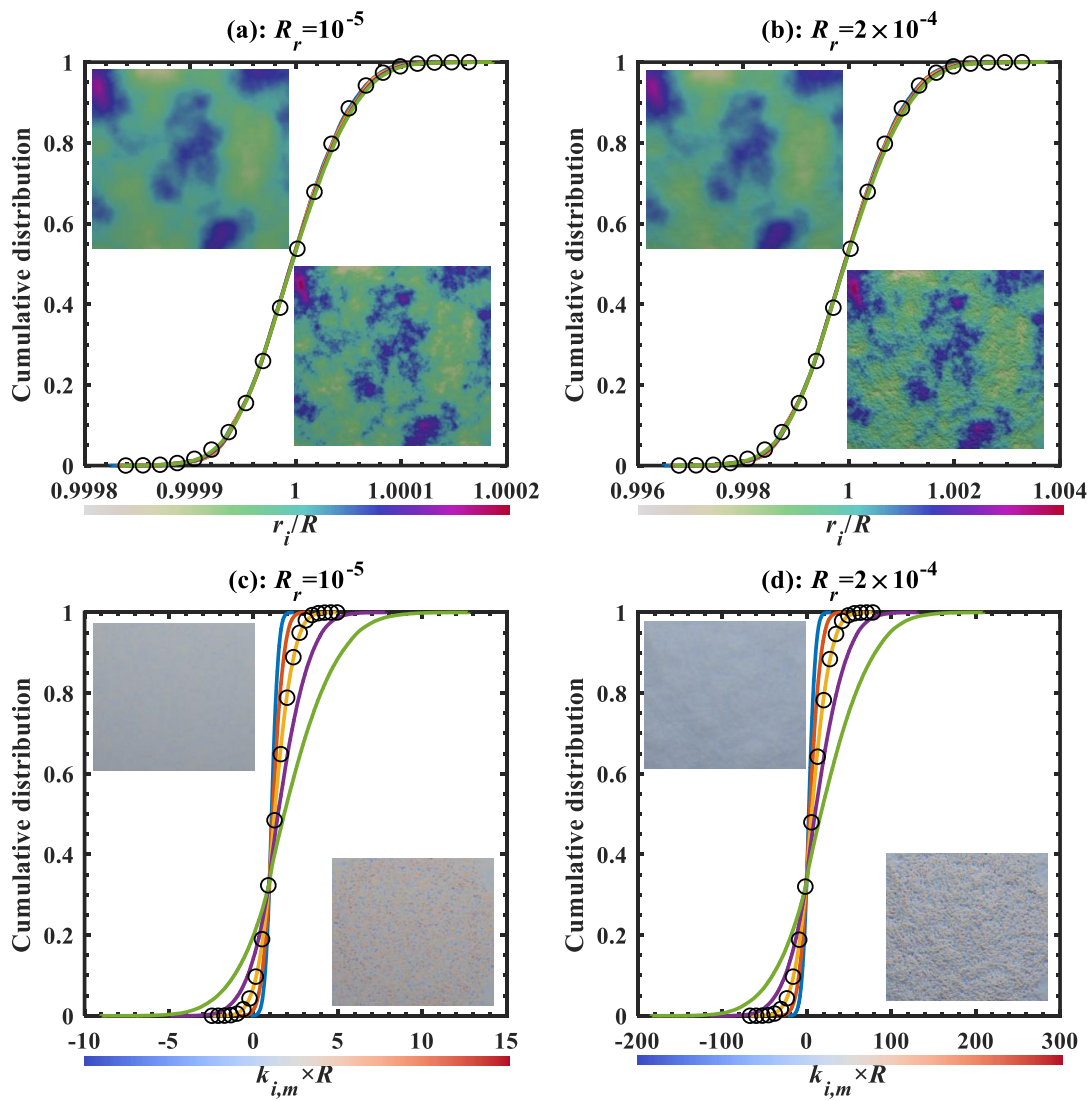

**Fig. 5** Cumulative distributions of radii and curvature values for $R_r$ values of $10^{-5}$ and $2\times10^{-4}$. Blue, orange, yellow, purple and green curves represent *FD* vales of 2.1, 2.2, 2.3, 2.4 and 2.5; Black circles represent the fitted Normal distributions of *FD*=2.3.

### 2.2 FEM model

In this study, a surface mesh was firstly generated via a MATLAB script and was converted into a solid using an open-source mesh generator Gmsh (Geuzaine and Remacle, 2009).

The FEM simulations presented here were conducted using commercial finite element software, ABAQUS. According to the Hertzian contact model, the effective radius, $R^*$, is defined as $1/R^* = 1/R_1+1/R_2$, where $R_1$ and $R_2$ are radii of contacting spheres. The effective contact modulus $E^*$ is defined as $1/E^* = (1-v_1^2)/E_1+(1-v_2^2)/E_2$, where $E_1$, $E_2$ and $v_1$, $v_2$ are the elastic Young's moduli and Poisson's ratios of the corresponding parts. The contact radius $a$ can be calculated by $a = \sqrt{R^*\delta}$, where $\delta$ is the indentation depth (Johnson, 1985). To obtain sufficient accuracy in simulating Hertzian contact and efficiency in FEM-based contact behaviours, extremely fine meshe sizes of $0.0015R$, were used within the potential contact region, as illustrated by the inner yellow circle in Fig. 2. The vertical height between the boundary mesh-fine region and the lowest point was $0.0225R$, with $R$ being the radius of the sphere. In the transition zone (between the yellow and red circles in Fig. 2), the average mesh size was around $0.1R$. Other parts were meshed roughly with a mesh size of $0.4R$. Following Section 2.1, polar coordinates $(\theta, \varphi)$ of spherical mesh vertices were applied in Eq. (17) for rough spheres with given $FD$ and $R_r$. The corresponding triangulated surface was then directly implemented as FEM surface mesh.

By means of regionally varying mesh resolutions, solid elements of high mesh qualities (e.g. $0.5 < h/l < 1.5$, where $h$ is the height, and $l$ is the length) can be obtained, by converting the triangular surface mesh to a 4-node tetrahedral volume mesh (C3D4). The rigid platen was composed of 4-node rigid quadrilateral elements (R3D4), with mesh size equalling to $0.0015R$. Jackson and Green (2005) have pointed out that linear FEM elements can yield the same result as that of quadratic elements. Concurrently, compared with hexahedron tetrahedron was chosen due to the efficiency of triangles in depicting complicated rough surfaces. The values of $v$ and friction coefficient were both set to 0 to isolate the purely normal contact response (Borri-Brunetto et al., 2001; Pei et al., 2005; Hyun and Robbins, 2007). Indeed, Hyun et al. (2004) suggested that the value of Poisson's ratio has little effect on the relationship between normal contact area and force.

Elements of the coarsest mesh size of the semi sphere were fixed in all directions, while the movements of finer elements were not constrained. The rigid platen moves only vertically, with an increasing normal displacement load to compress the sphere.

FEM simulations were conducted in Abaqus Explicit environment, only key information is given here with more details available in the Abaqus Users' manual (2016). The constitutive elastic law of bulk material was assumed to be isotopically linear in terms of first-order tetrahedron (C3D4) elements (Hyun et al., 2004; Abaqus, 2016). Full integration was considered for calculating the virtual work. The total mass of each element was defined by lumped mass matrix and averagely distributed over its four nodes. Also, a four-pointed integration scheme, where distributed loads were integrated with three points, was applied. An explicit integration scheme with an augmented Lagrangian framework was considered,

$$\boldsymbol{M} \cdot \ddot{\boldsymbol{u}} + \boldsymbol{F}_i - \boldsymbol{F}_e = \boldsymbol{0}, \tag{19}$$

where $\ddot{\boldsymbol{u}}$ is the acceleration vector, $\boldsymbol{M}$ is the diagonal mass matrix and $\boldsymbol{F}_i$ and $\boldsymbol{F}_e$ are internal and external force vectors. The central difference integration framework is implemented to discrete Eq. (19) in time:

$$\dot{\boldsymbol{u}}_{n+1/2} = \dot{\boldsymbol{u}}_{n-1/2} + \frac{\Delta t_{n+1} + \Delta t_n}{2} \ddot{\boldsymbol{u}}_n, \tag{20}$$

$$\boldsymbol{u}_{n+1} = \boldsymbol{u}_n + \Delta t_{n+1} \dot{\boldsymbol{u}}_{n+1/2}, \tag{21}$$

$$\Delta t = \sqrt{\frac{\rho}{E}} L_{min}, \tag{22}$$

where $u$ is a freedom degree, $n$ means $n$-th time step or increment, and $\Delta t$ denotes time step; $\rho$ is the density of the bulk material, $E$ is the elastic modulus, and $L_{min}$ is the minimum length of mesh size.

## 3. Results and Discussion

Here we present simulation results from 25 contacts with rough spheres and one with a smooth sphere, to elucidate the influences of $R_r$ and $FD$ on the resulting contact pressure, contact area and contact stiffness.

### 3.1 Contact pressure

Maps of the contact pressure (corresponding to the sum of nodal contact forces divided by the associated element face area) are illustrated in Fig. 6. Applying a rigid flat platen as a counter surface and the absence of friction results in nodal forces that are all perpendicular to the contact plane. The results suggest that greater fractality, as indicated by larger $FD$ values, results in a more heterogenous pressure contour, with larger maximum asperity contact stress. An increasing $R_r$ resulted in an evident drop in the total contact area, with increasing stress concentration, and larger maximum radial distance.

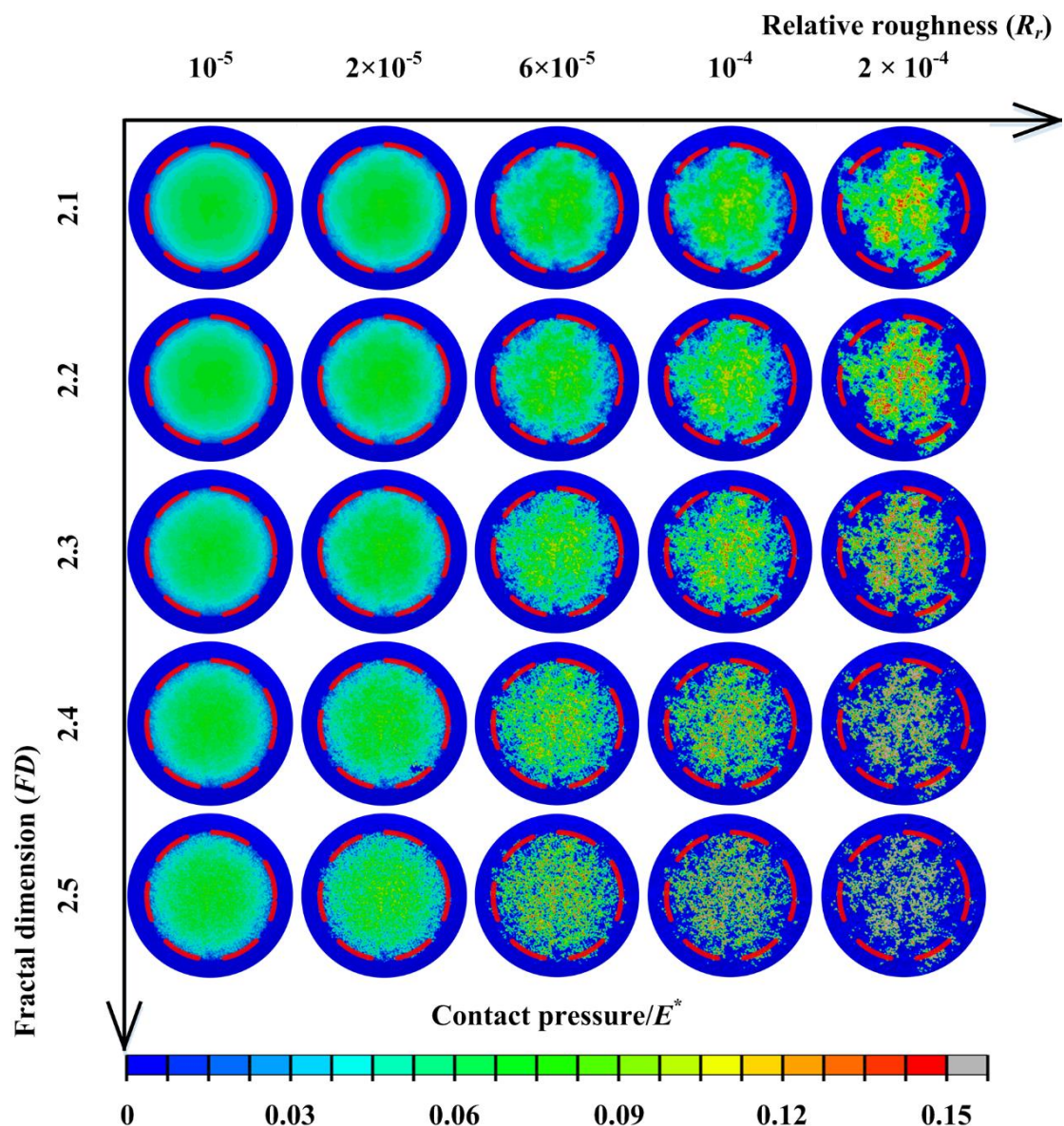

**Fig. 6** Normalized contact pressure ($p/E^*$) simulated using a constant compressive force ($F/ER^2 = \frac{4}{3} \times 0.1^3$) for varied conditions of $R_r$ and $FD$ (red circles represent the Hertzian contact boundaries for the same normal force).

To quantify the influence of surface morphology on contact stress distributions, contact pressure at varying radial distances ($a$) on different regions was further calculated by the ratio of nodal contact force to contact area from square R3D4 elements (Fig. 7):

$$p(a) = \frac{\sum_i F_i}{\pi(a+d)^2 - \pi d^2}, \qquad (23)$$

where $F_i$ is the $i$-th nodal contact force on the rigid flat platen and $d$ is the width of the circular ring. Each individual ring was divided into several subdomains having identical angles (shown by $\alpha$ in Fig. 7) for obtaining further information regarding the angular pressure distribution. As is shown in Fig. 7, the mean pressure in blue, yellow and red sub-rings were calculated from the corresponding nodes within these domains.

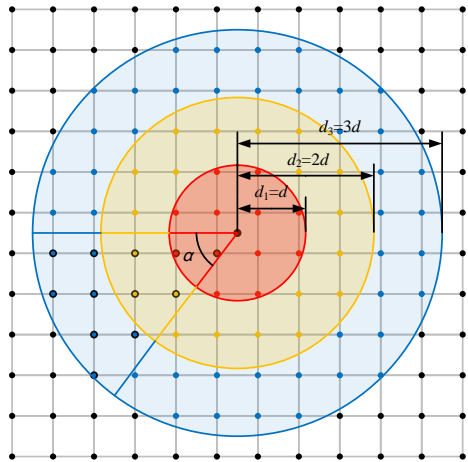

**Fig. 7** Schematic for evaluating local contact pressure.

The radial stress distribution for various values of $R_r$ and $FD$ are presented in Fig. 8. A nearly complete coincidence is found between the Hertzian solution and results for $R_r = 0$ in Fig. 8 (g), validating the simulation framework presented in this study. As shown in Fig. 8 (a), the stress distribution for lower values of $R_r$ deviates from the Hertzian solution. A larger fluctuation in contact pressure can be generally observed for larger $FD$. It is interesting to note that at large values of $R_r$ there is a contact stress peak at the position of $a/R \approx 0.05$, due to the randomness of particle morphology which results in a fall of heights of particle surface asperities, as in Fig. 5 (a) and (b). In Fig. 8, at $a/R \approx 0.05$, the maximal values of contact pressure exhibit a sharp increase while the minimal values tend to be smoother, demonstrating the significant influence of local surface features on contact pressure. Compared with deterministic models, such as Greenwood and Tripp (1967), our numerical results here are more consistent with optical experiments (Sharp et al., 2018) where the maximum normalized contact pressure (at $a/R=0$) was found to be greater than that of the Hertzian solution.

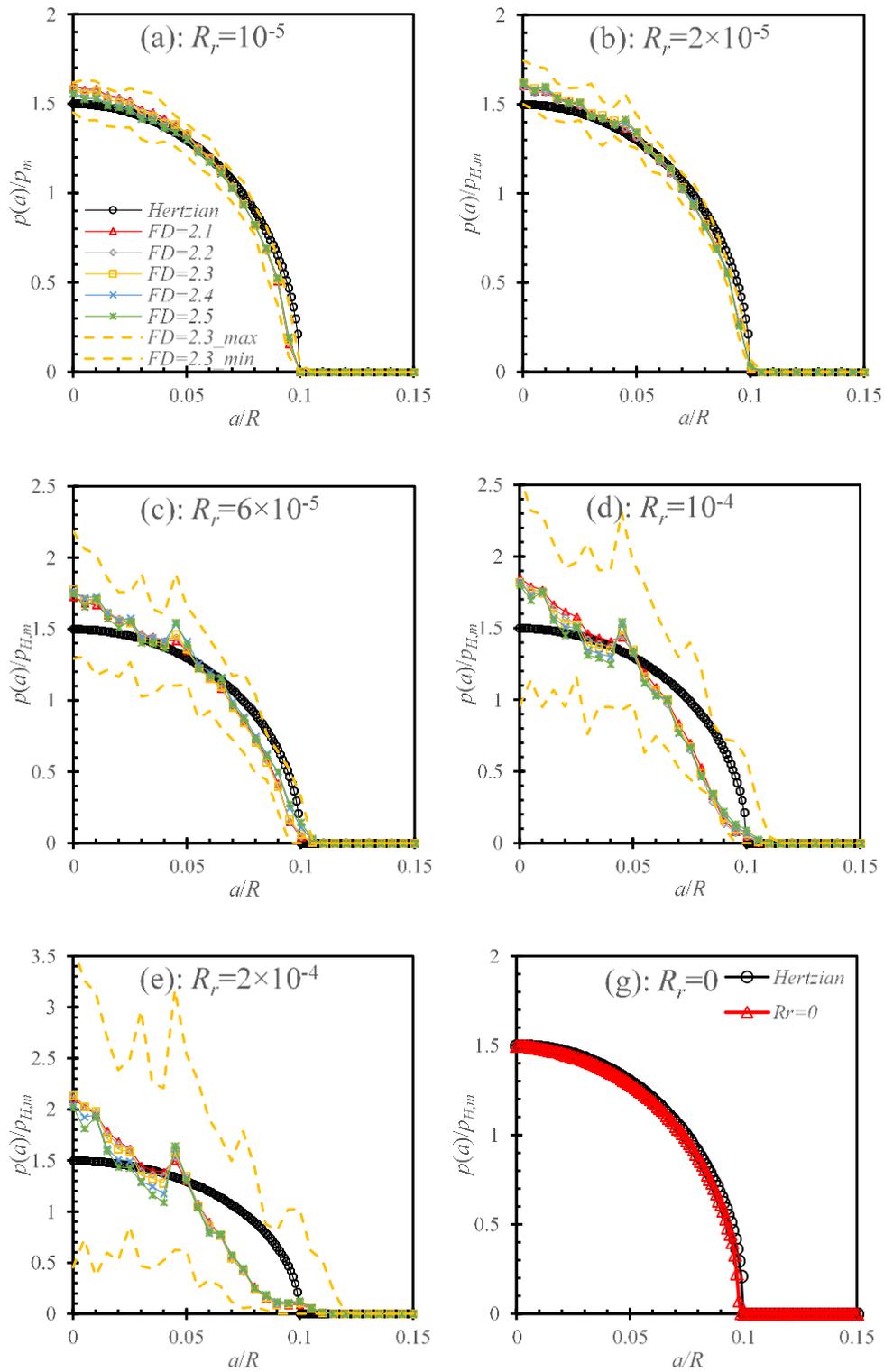

**Fig. 8** Normalised contact pressure at varying radial distance for different values of $R_r$ and *FD*.

## 3.2 **Contact area**

In this study, contact area (*A*) was calculated based on the sum of the deformable element face areas with nodes contacting the rigid platen. According to the Hertzian contact model, the contact radius of a smooth sphere compressed against a rigid flat is given as

$$r_H = \left(\frac{3FR}{4E^*}\right)^{\frac{1}{3}}. \tag{24}$$

Here, the contact area is normalised by $\pi R^2$ and the force is normalised by $E^*\pi R^2$,

$$\frac{A}{\pi R^2} = \left(\frac{3\pi}{4}\right)^{\frac{2}{3}} \times \left(\frac{F}{E^*\pi R^2}\right)^{\frac{2}{3}}. \tag{25}$$

As shown in Fig. 9, the relationship between normalized force and area for small $R_r$ was found to agree with the Hertzian solution. With increasing $FD$ and $R_r$, $A/(\pi R^2)$ was found to deviate more significantly from the Hertzian solution. Fig. 10 illustrates the fractal dimension of the contact boundary, $FD_{BC}$, calculated using a box-counting method, for varied $FD$ and $R_r$ at a constant compressive force. For lower $R_r$ values the outlines are tortuous to a similar extent with the increase of $FD$, meanwhile for higher $R_r$ the $FD$ of contact boundary increases.

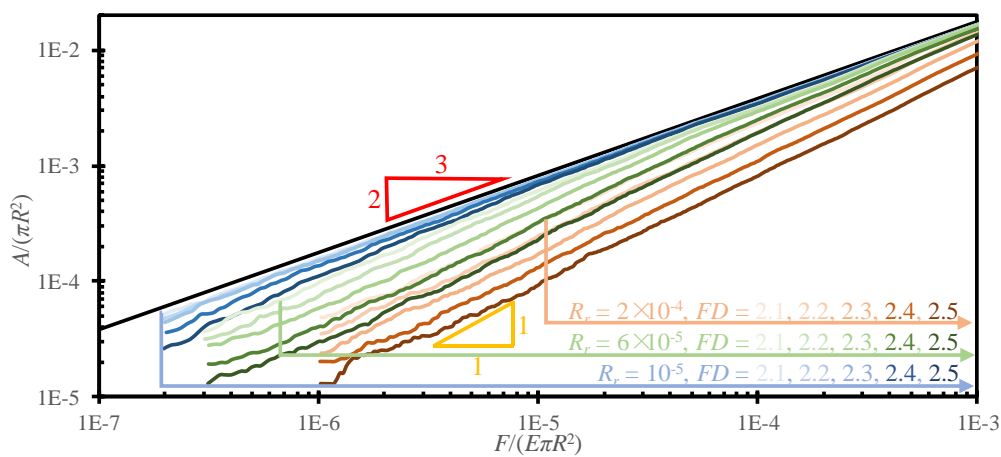

**Fig. 9** Contact area ($A/(\pi R^2)$) V.S. compression force ($F/(E\pi R^2)$). Hertzian solution given by the solid black line.

In the GT model (Greenwood and Tripp, 1967), Hertzian contact becomes applicable only when $F > N_2$, with

$$N_2 = 100 N_1 = 100 S_q^* E^* \sqrt{2R^* S_q^*}, \tag{26}$$

where

$$S_q^* = \sqrt{S_{q1}^2 + S_{q2}^2}, \tag{27}$$

where $S_q$ is the root mean square (RMS) surface roughness and can be denoted by the standard deviation of the normal distribution to fit asperity 'radius' distribution. We apply a power law, i.e., $\frac{A}{\pi R^2} = \alpha(\frac{F}{E^*\pi R^2})^\beta$, to correlate normalized forces with contact area for contact forces smaller than $N_2$. For all 25 cases, values of goodness of fit (calculated as R-square) were more than 0.99. As shown in Fig. 11 (a) and (b), for the lowest relative roughness (e.g. $R_r = 10^{-5}$) the values of both $\alpha$ and $\beta$ do approach closely to those of the Hertzian solution in Eq. (25) ($\left(\frac{3\pi}{4}\right)^{\frac{2}{3}}$ and $\frac{2}{3}$, respectively), however for higher $R_r$ both of them are not well described by the Hertzian solution and more fluctuations appear, although convergence to the Hertzian solution occurs at high loads as the limit of linear elastic material is achieved, thus the contact forces cannot be enlarged for the rationality of pure elasticity in FEM models. By contrast, for spheres of both high- and low-roughness, Pastewka and Robbins (2016) showed the convergence of the contact area to the Hertzian solution at extremely high loads. This discrepancy may be attributed to: (i) Sphere rigidity limiting the study of deformation-induced contact (Li et al., 2018); or (ii) plasticity in molecular simulations, which significantly influences contact behaviour (Song et al., 2016, 2017.), herein just purely elastic contact is studied.

Contact island distribution plays a vital role in thermal and electrical conductivity properties at the interfacial scale in granular materials (Persson et al., 2010; Zhai et al., 2016b). As shown in Fig. 6, larger values of $R_r$ and $FD$ bring about more scattered distributions of contact stress. Increasing loads will result in larger contact areas by expanding existing contact islands and the formation of new contact islands, i.e., fringing the boundary of contact zone. This process is also accompanied by the merging of existing contact islands. In order to quantitatively evaluate merging and fringing processes of contact islands, we implemented image segmentation techniques to separate connecting contact islands, shown in Fig. 12. The quadrilateral R3D4 elements composing the rigid flat were considered as image pixels with greyscale values denoting the nodal contact forces. Pixels corresponding to the maximum normal contact force are set to a greyscale value of 255, with a linear representation of contact forces. A contact stress contour map, in which values denote nodal contact stresses, is drawn. Then, a threshold value equal to 0.3 times the maximum nodal contact stress is chosen to separate contact islands. Finally, the number of contacting regions forming separated contact islands can be obtained. Furthermore, besides the case in Fig. 6, 3 more cases of $R_r$ and $FD$ equal to $6\times10^{-5}$ and 2.3, while from different sets of SH coefficients, were added for more universally statistical features. In summary, Fig. 13 (a) provides the probability density function (PDF) of segmented contact island area with gradually enlarged contact forces for 4 cases of $R_r = 6\times10^{-5}$ and $FD = 2.3$. The merging of centre islands and fringing the boundary demonstrate competitive tendency. Notably, for smaller islands ($A_c/(\pi R^2) < 1.5\times10^{-5}$), the PDF seems to have a consistent power-law distribution. Simultaneously, as shown in Fig. 13 (b), obtained contacts for varying loads contain microcontacts, of which areas nearly conform to the same Weibull distribution over a wide range of self-similar length scales. By assigning the same Weibull modulus of 0.503 for smaller contacting islands ($Ln(A_c/A_{c,0}) < 3.5$), the goodness-of-fitting coefficients (R-square values) for five consequent loading stage range from 0.951 to 0.997. It is reasonable to postulate that the Weibull modulus is correlated to the fractal nature of contacting surfaces, warranting future systematic studies focusing on the variations of *FD* and $R_r$.

For rough interfaces, contact area distributions can have significant implications. For interfacial electrical conduction of granular materials, for example, these microcontacts of different sizes transport electrical current through different conduction mechanisms (Zhai et al., 2015), including Holm contacts, Sharvin contacts and electron tunnelling. When the size of contacting asperities is comparable or smaller than the average electron mean free path, electrons travel ballistically across the microcontacts (Zhai et al., 2016b). The herein proposed power-law correlation for describing the contact area distribution and its evolution enables the identification of the dominant conduction mechanisms across multiple length scales.

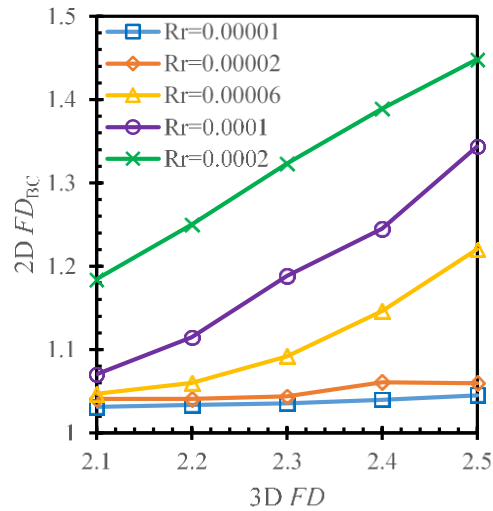

**Fig. 10** Relationship between the fractal dimension of contact area, $FD_{BC}$, calculated by a 2D box-counting method and *FD* of particle surface for various $R_r$ at the same compressive force ($\frac{F}{E^*\pi R^2} = \frac{4}{3} \times 0.1^3$) in Fig. 6.

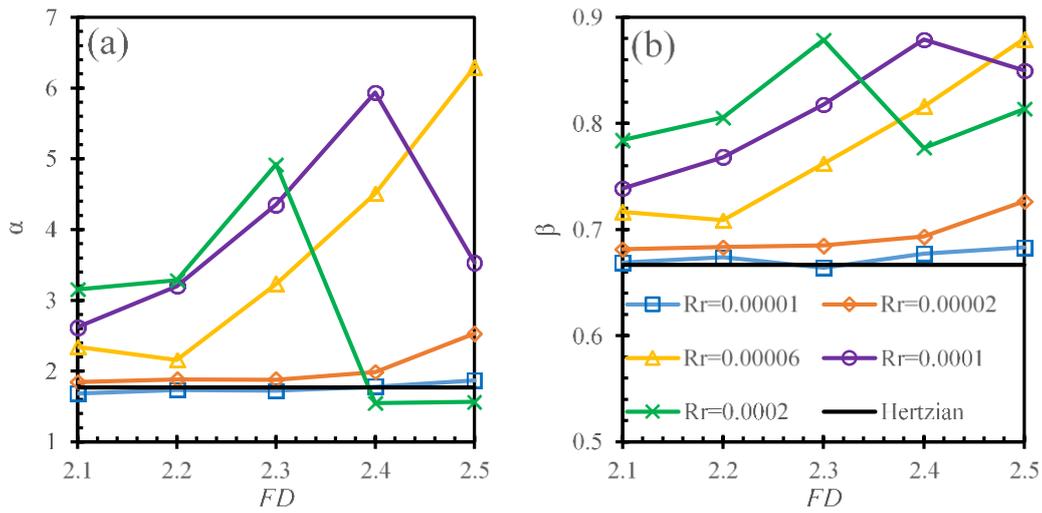

**Fig. 11** Relationships between *FD* and $\alpha$ and $\beta$ for various $R_r$.

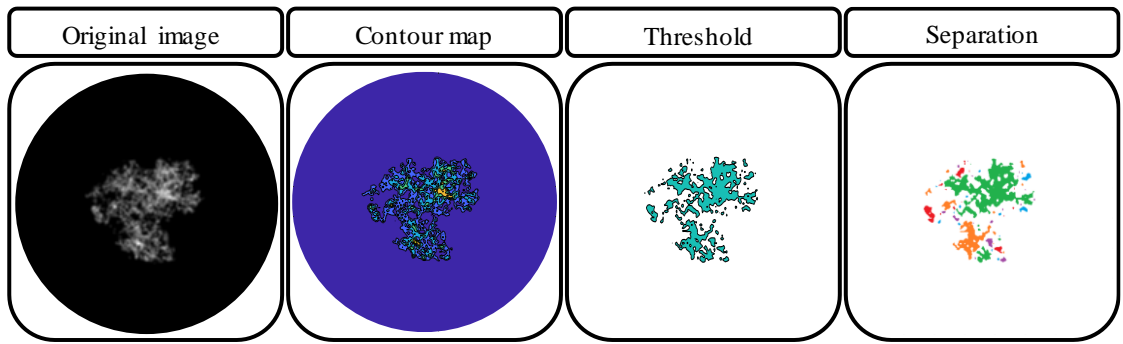

**Fig. 12** Segmentation process of contact islands.

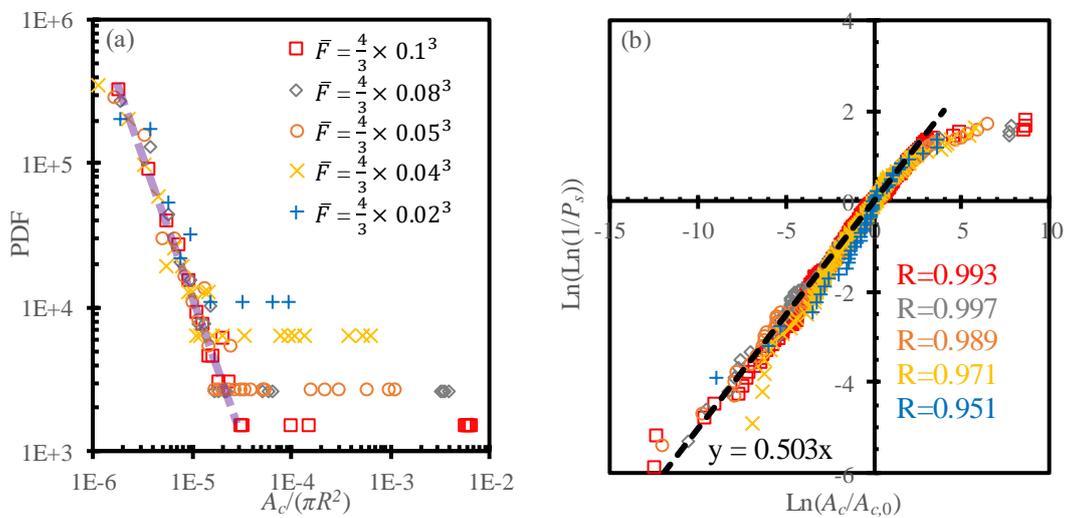

**Fig. 13** Distributions of normalized contact island area ($A_c$) for various normalized contact forces $\bar{F}$: (a) Probability density distributions (PDFs); (b) Weibull distributions. $R_r$ and $FD$ equal to $6\times10^{-5}$ and 2.3, respectively; The transport purple dot line is just for illustration purpose, where the relations between $\bar{F}$ and PDF at the decrease period are roughly linear; $A_{c,0}$ is the characteristic area where 37 % of the contact regions survive.

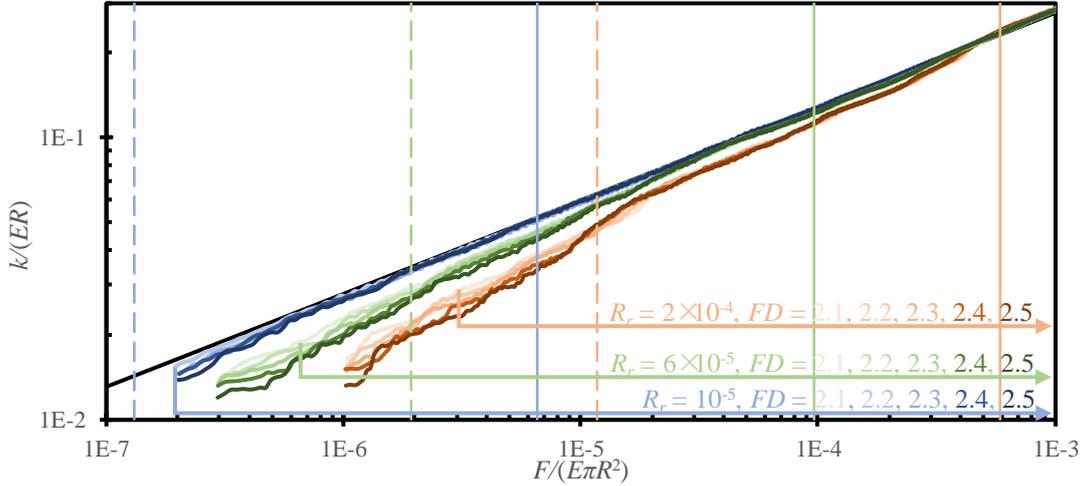

**Fig. 14** Dimensionless normal contact stiffness ($k/(ER)$) vs. compressive force ($F/(E\pi R^2)$). Hertzian solution was given by the solid black line. The red solid and vertical lines denote values of $N_1$ and $50N_1$ (or $0.5N_2$), respectively.

### 3.3 Normal contact stiffness

Normal contact stiffness is calculated by the following equation:

$$k_i = \frac{\Delta F_i}{\Delta d_i} = \frac{F_{i+1}-F_i}{d_{i+1}-d_i}, \tag{28}$$

where subscript $i$ indicates the $i$-th time step in the explicit FEM scheme; $k_i$ is the normal contact stiffness; $F_i$ and $d_i$ represent respectively the normal force acting on the rigid plate and its displacement. In DEM, it is widely accepted (Greenwood and Tripp, 1967; Yimsiri and Soga, 2000; Pohrt and Popov, 2013) to consider contact stiffness based on Hertzian contact solutions when the overall contact force is large enough, i.e., $>N_2$. The relation between contact stiffness and force in the Hertzian contact model is (Johnson, 1985):

$$\frac{k}{E^*} = \left(6R^* \frac{F}{E^*}\right)^{\frac{1}{3}}. \tag{29}$$

Here, the contact stiffness $k$ is normalized by $ER$, and Eq. (29) can be rewritten as:

$$\frac{k}{ER} = (6\pi)^{\frac{1}{3}} \times \left(\frac{F}{E\pi R^2}\right)^{\frac{1}{3}}. \tag{30}$$

The power law $\frac{k}{ER} = \alpha(\frac{F}{E^*\pi R^2})^\beta$ was employed here to fit the relationship between normalized contact stiffness and forces. We consider two segments, $F < N_1$ and $F > 50 N_1 = 0.5\, N_2$.

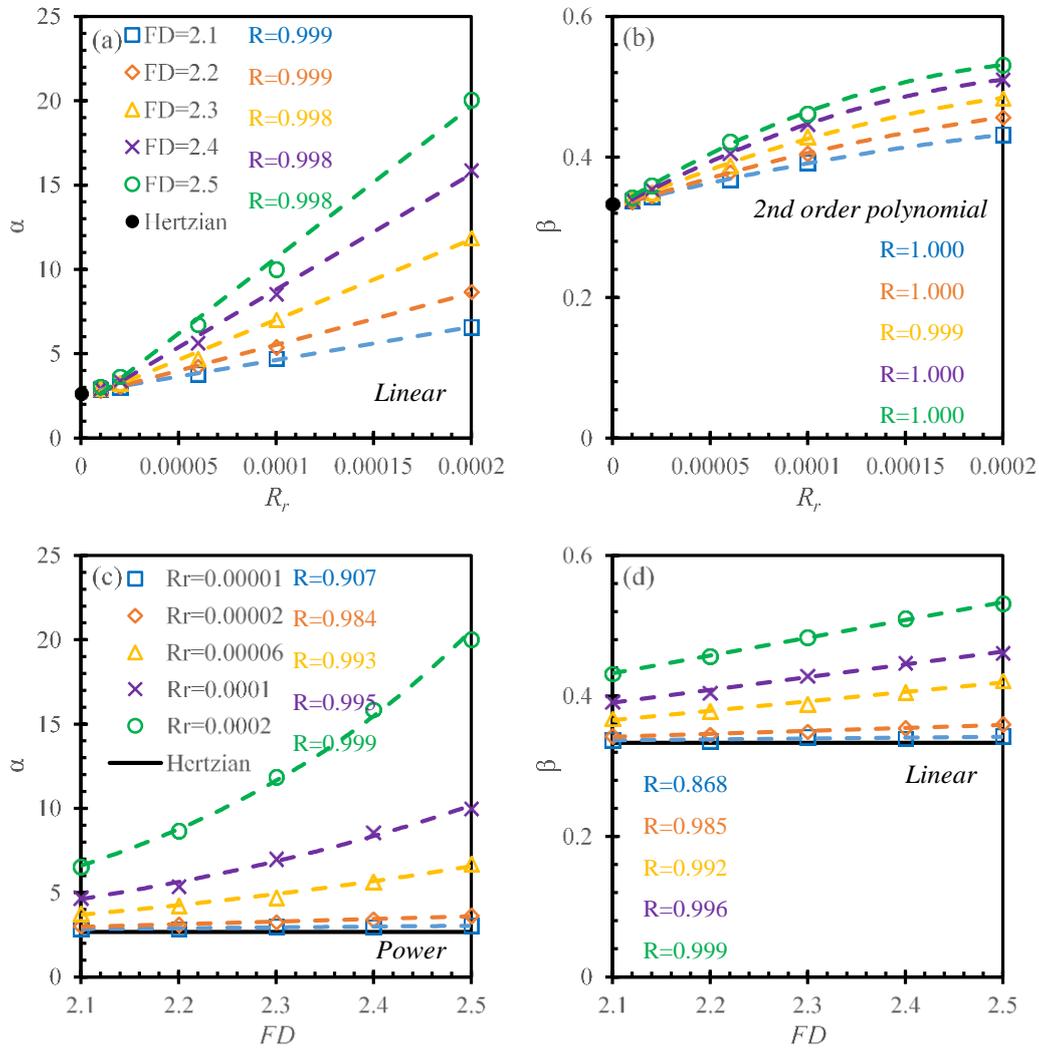

**Fig. 15** Relationship between fitting parameters of $k$-$F$ curves and $R_r$ ((a) and (b)) and $FD$ ((c) and (d)) for $F < N_1$.

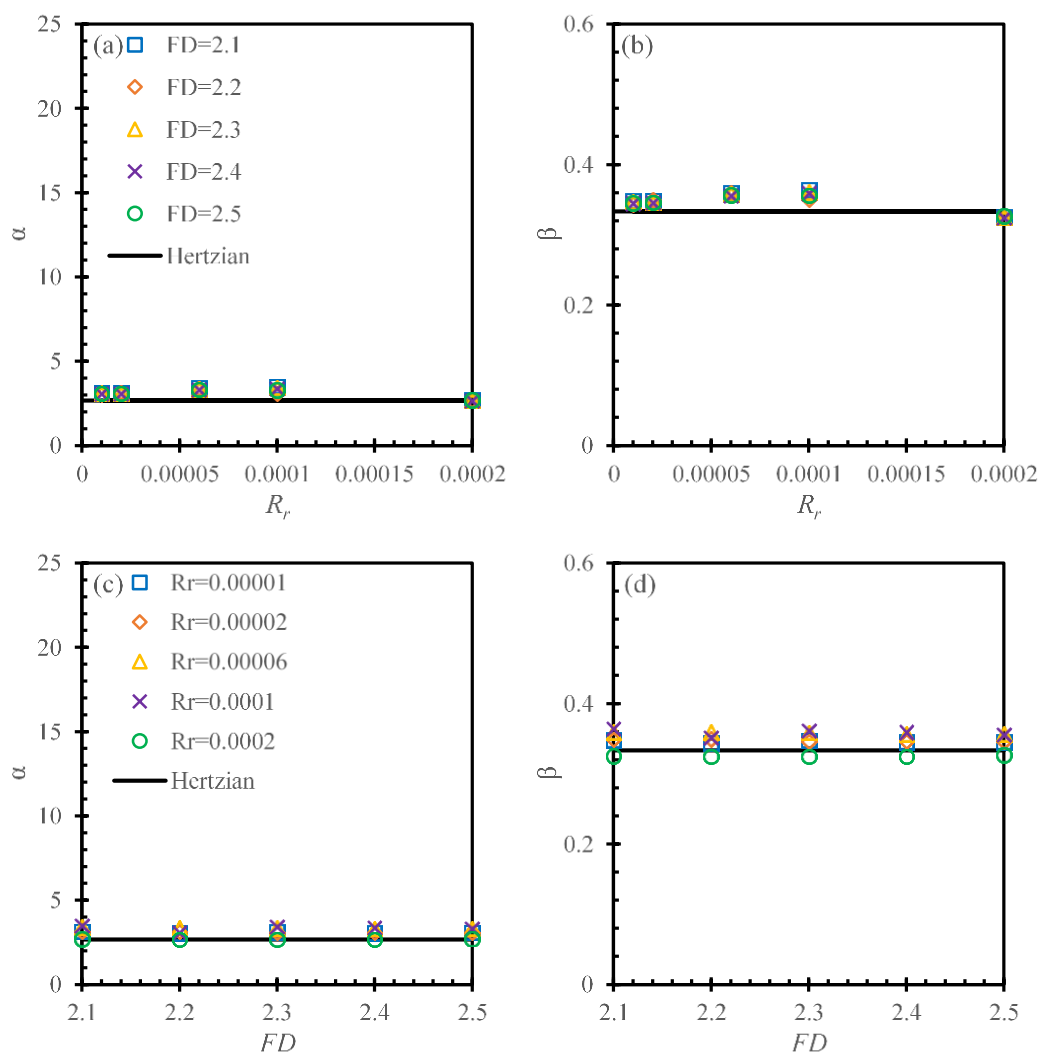

**Fig. 16** Relationships between fitting parameters of *k-F* curves and $R_r$ ((a) and (b)) and *FD* ((c) and (d)) for $F > 0.5N_2$.

Figs. 15 and 16 illustrate the relationships between fitting parameters and relative roughness and fractal dimension, respectively. For cases of $>N_2$, the Hertzian contact solution is considered applicable for the calculation of contact stiffness. Meanwhile for cases of $<N_1$, both fitting parameters are found to be influenced by $R_r$ and *FD*. It further reveals that the application of RMS alone to describe a rough-sphere contact is insufficient, and correlations between 'radii', evaluated by fractal dimension, should be included. Based on the presented parametric studies, one simple empirical model is proposed for determining the dependence of *k* on *F* for contacts of rough spheres:

$$\frac{k}{E^* R^*} = \left(\sqrt[3]{6\pi} + R_r \cdot D_f^{4\pi}\right) \times \left(\frac{F}{E^* \pi R^{*2}}\right)^{\frac{1}{3} + 351 R_r \cdot D_f}, \tag{31}$$

where $D_f$ denotes *FD*. Note this proposed correlation can be reduced to Hertzian contact while $D_f$ or $R_r$ approaches zero, i.e., cases of smooth spheres.

Fig. 17 represents the 3D plots describing $\alpha$ and $\beta$ versus $FD$ and $R_r$, showing a reasonable goodness-of-fitting for all simulation cases presented in this work. Correlations developed here can be extended to incorporate the multiscale features in contact mechanics of rough surfaces, and readily to be implemented in other numerical schemes where the inter-particle contact model is pivotal, such as Discrete Element Methods (DEM), under a consequent multiscale modelling scheme. Moreover, for applications involving contact rough interfaces, the effects of hierarchical properties of the surface structure can be estimated using the above proposed correlation, in conjunction of measuring surface profiles.

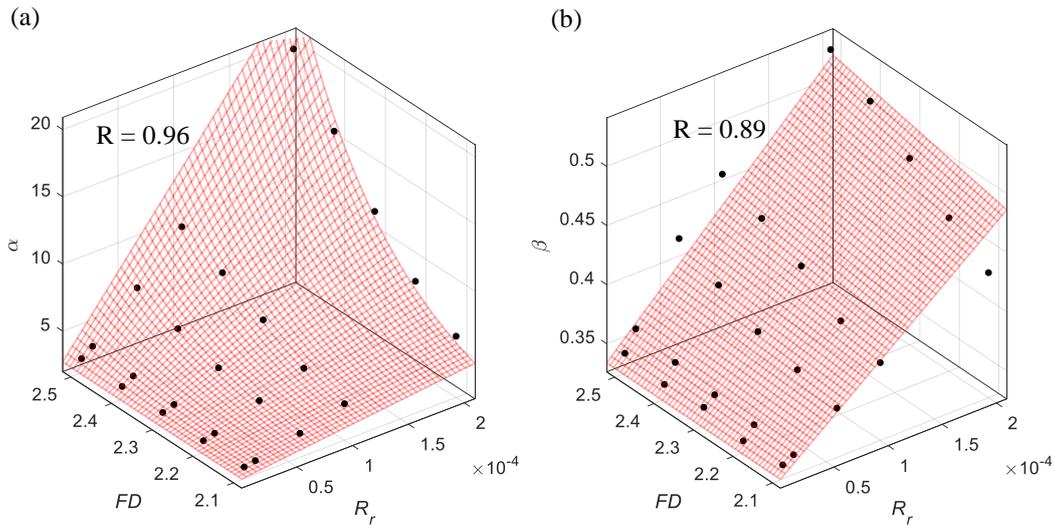

**Fig. 17** Unique power law parameters ($\alpha$ in (a) and $\beta$ in (b)) to denote normal contact stiffness of rough sphere contact in $\alpha$ and $\beta$ -$FD$-$R_r$ spaces.

## 4. Discussion and Conclusion

It is reasonable to compare the contact response from rough spheres and nominally flat rough surfaces at low loads, as a small section of a sphere's surface can be seen as a rough platen under such conditions. Some reports indicate that contact responses of rough sphere approach the results of nominally flat surfaces (Pohrt an Popov, 2013). However, mapping a sphere surface to a flat without geometrical distortion is not possible. Thus, it is difficult to quantitatively compare normal contact response between them, bridging such responses with roughness (e.g., fractal dimension, root mean square roughness and gradient, and high and low wavelength). Despite these differences, at low loads (e.g. $F < N_1$ in GT model) our model states that $k \propto F^{D_f}$, which is consistent with BEM simulations of contact behaviour of nominally flat surface in Pohrt and Popov (2012):

$$\frac{k}{E^*\sqrt{A_0}} = \frac{\pi D_f}{10}\left(\frac{F}{E^* h\sqrt{A_0}}\right)^{0.2567 D_f}. \tag{32}$$

where $h$ is root mean square roughness, $D_f$ is the fractal dimension determined by PSD, and $A_0$ is the projection area of nominally flat platen. Compared with Eq. (31), both models have two fitted or approximated parameters, which are $\frac{\pi D_f}{10}$ and $0.2567 D_f$ in nominally flat platen and $R_r \cdot D_f^{4\pi}$ and $351 R_r \cdot D_f$ for the case of a rough sphere, considering the influences of both roughness and fractal dimension. Furthermore, two ideal conditions are covered, where the stiffness of a rigid flat is infinite while a smooth sphere conforms to the Hertzian solution.

Besides contact stiffness, it is necessary to mention the most accepted relation between contact area and force. At low loads, from the analytical solutions of contact behaviour of nominally flat surfaces with spherical (Greenwood and Williamson, 1966) or ellipsoidal asperities (Bush et al., 1975), the relation between contact area and force is linear, which is consistent with FEM simulations including irregular asperities considering plasticity (Song et al., 2016, 2017) or not (Hyun et al., 2004). As Fig. 9, this relation of spherical surfaces with higher $R_r$ and $FD$ approximates more linear, and the influence of $FD$ is not evidently weaker than that of $R_r$, because rougher sphere behaves more like nominally rough flat platen in small scale of surface.

In this study we propose an effective framework for generating realistic fractal rough particle surfaces and the corresponding FEM meshes, based on spherical harmonics (SH). The effect of local asperity curvature is incorporated explicitly in this study, via the term of relative roughness, the ratio between global curvature and local roughness. Compared to the classical contact model (e.g. GT model), the local asperity curvature (or radius) can be included by using ultra-high SH resolution to capture multiscale morphological features, namely the local curvature and roughness. Finite element analyses of rough sphere contacts demonstrated the morphological dependencies of contact behaviour, e.g., contact area and contact stiffness, with a focus on relative roughness and fractal dimension of surface features. The main findings and conclusions are summarized as follows:

- Contact behaviour depends strongly on surface features, characterised here by relative roughness and fractal dimension, in the regime of relatively low loads.
- With increasing contact pressure, competition exists between contact area merging and the formation of new small contact islands occurring around contact area boundaries for spheres of intermediate roughness. For spheres with highly rough surfaces, contact islands do not tend to merge, and increasing contact pressure is associated mainly with the formation of new contact islands. During the contact, individual contact islands evolve and merge, which follows a Weibull-type distribution independent of the loading level.
- For relatively small contact forces, the normal contact stiffness of rough spheres is dependent on both fractal dimension and relative roughness and is well-described by power law correlations, differing from Hertzian theory at low load. The contact stiffness presents a power law behaviour with the applied force, Empirical relations between contact stiffness and load have been proposed considering topological indices, including the relative roughness and fractal dimensions.

The numerical framework presented here for the study of contact mechanics of rough particles warrants further investigation. More specifically, using the developed SH-mesh based FEM schematic in this study, the simulation of contact behaviours of rough particle with globally irregular shapes can be conducted. In addition to the normal contact behaviours presented here, further studies are required to elucidate relationships between particle friction behaviour and asperity level morphology.

**Acknowledgements**

The authors would like to acknowledge the support from The University of Sydney through SOAR Fellowship and the Australian Research Council through its Discovery Project DP170104192.

**Appendix I: Six kinds of particles and their high-degree SH descriptors**

This part mainly demonstrates the efficiency of SH-based fractal dimension (*FD*) to characterize particle morphology. Although in the previous study (Wei et al., 2018), the linear relation between SH descriptor and degree *n* in log-log scales has been illustrated, only two kinds of sand particles were contained, and they are of the similar size (e.g. the equivalent-volume-sphere is from 0.5 mm to 2 mm). For further improving, six kinds of particles (Bullard, 2014), as shown in Fig. 3, in concrete and with size laying in larger scopes (e.g. from sands in motor to aggerate) from Virtual Cement and Concrete Testing Laboratory (VCCTL) are covered.

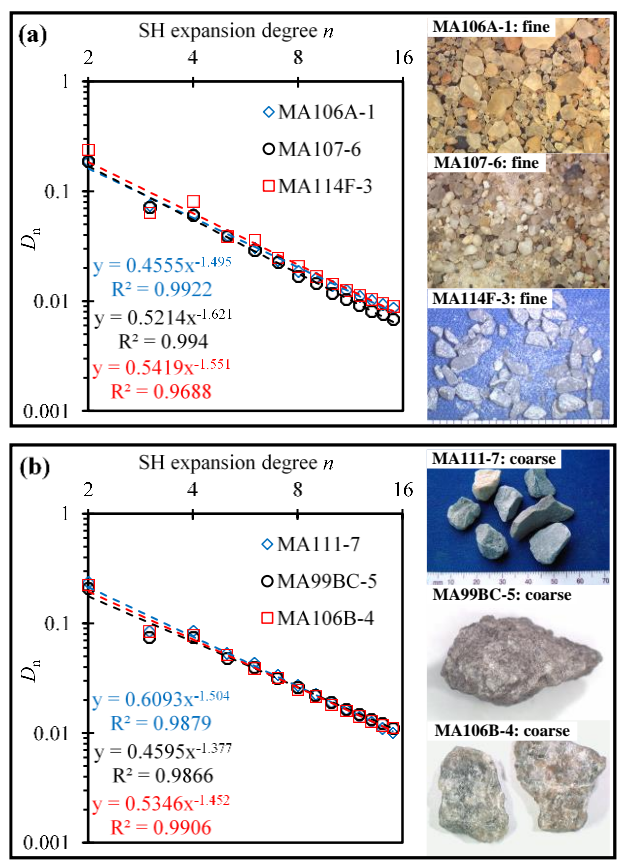

Fig. A-1 The relations between SH descriptor $D_n$ and SH expansion degree for six kinds of granular materials used in concrete (adopted from Bullard (2014)): (a): Fine sands; (b): Coarse aggregates.

**Appendix II: Deduction of mean curvature H**

Local surface properties (e.g., maximum curvature and minimum curvature) are evaluated by differentials of the surface equation. The mean curvature value of a 3D surface point is the average of its two principal curvature values. We begin from the normal vector ($\hat{n}$) of the SH-based particle surface:

$$\hat{n} = \frac{\vec{X_\theta} \times \vec{X_\varphi}}{|\vec{X_\theta} \times \vec{X_\varphi}|}, \tag{A-1}$$

where $\vec{X} = (x, y, z)$ is the surface vector. The partial differentiations respecting to polar coordinates write:

$$X_\theta = \frac{\partial X(\theta,\varphi)}{\partial \theta} = \sum_{n=1}^{2000} \sum_{m=-n}^{n} c_n^m \frac{\partial Y_n^m(\theta,\varphi)}{\partial \theta}, \tag{A-2}$$

$$X_\varphi = \frac{\partial X(\theta,\varphi)}{\partial \varphi} = \sum_{n=1}^{2000} \sum_{m=-n}^{n} c_n^m \frac{\partial Y_n^m(\theta,\varphi)}{\partial \varphi}. \tag{A-3}$$

According to Koenderink (1990), coefficients of the first (*I*) and the second (*II*)

fundamental forms of surface vector are related to surface curvatures and they are defined as

$$I = d\vec{X} \cdot d\vec{X}$$

$$I = Ed\theta^2 + 2Fd\theta d\varphi + Gd\varphi^2. \tag{A-4}$$

where

$$E = \vec{X_\theta} \cdot \vec{X_\theta}$$

$$F = \vec{X_\theta} \cdot \vec{X_\varphi}$$

$$G = \vec{X_\varphi} \cdot \vec{X_\varphi}; \tag{A-5}$$

and

$$II = d\vec{X} \cdot d\hat{n}$$

$$II = Ed\theta^2 + 2Fd\theta d\varphi + Gd\varphi^2. \tag{A-6}$$

where

$$L = -\vec{X_\theta} \cdot \hat{n}_\varphi$$

$$M = \frac{(\vec{X_\theta} \cdot \hat{n}_\varphi + \vec{X_\varphi} \cdot \hat{n}_\theta)}{2}$$

$$N = -\vec{X_\varphi} \cdot \hat{n}_\varphi. \tag{A-7}$$

Then the mean curvature value is given:

$$H = \frac{EN + GL - 2FM}{2(EG - F^2)}. \tag{A-8}$$

Deheng Wei, Chongpu Zhai, Dorian Hanaor, Yixiang Gan    2020, Contact behaviour of simulated    rough spheres generated with spherical harmonics, International Journal of Solids and Structures, 193, 54-68

## References

Abaqus User's Manual, R. (2016). Dassault Systémes Simulia Corp. *Providence, RI*.

Akarapu, S., Sharp, T., & Robbins, M. O. (2011). Stiffness of contacts between rough surfaces. *Physical Review Letters*, *106*(20), 204301.

Andrade, J. E., Lim, K. W., Avila, C. F., & Vlahinić, I. (2012). Granular element method for computational particle mechanics. *Computer Methods in Applied Mechanics and Engineering*, *241*, 262-274.

Barber, J. R., & Ciavarella, M. (2000). Contact mechanics. *International Journal of solids and structures*, *37*(1-2), 29-43.

Barber, J. R. (2003). Bounds on the electrical resistance between contacting elastic rough bodies. *Proceedings of the royal society of London. Series A: mathematical, physical and engineering sciences*, *459*(2029), 53-66.

Barrett, P. J. (1980). The shape of rock particles, a critical review. *Sedimentology*, *27*(3), 291-303.

de Bono, J. P., & McDowell, G. R. (2018). On the packing and crushing of granular materials. *International Journal of Solids and Structures*, *In Press*.

Borri-Brunetto, M., Chiaia, B., & Ciavarella, M. (2001). Incipient sliding of rough surfaces in contact: a multiscale numerical analysis. *Computer methods in applied mechanics and engineering*, *190*(46-47), 6053-6073.

Bowman, E. T., Soga, K., & Drummond, W. (2001). Particle shape characterisation using Fourier descriptor analysis. *Geotechnique*, *51*(6), 545-554.

Brutsaert, W. (1975). A theory for local evaporation (or heat transfer) from rough and smooth surfaces at ground level. *Water resources research*, *11*(4), 543-550.

Bullard, J. W. (2014). *Virtual Cement and Concrete Testing Laboratory: Version 9.5 User Guide* (No. Special Publication (NIST SP)-1173).

Buzio, R., Boragno, C., Biscarini, F., De Mongeot, F. B., & Valbusa, U. (2003). The contact mechanics of fractal surfaces. *Nature materials*, *2*(4), 233.

Cavarretta, I., Coop, M., & O'SULLIVAN, C. (2010). The influence of particle characteristics on the behaviour of coarse grained soils. *Géotechnique*, *60*(6), 413-423.

Chang, C. S., & Hicher, P. Y. (2005). An elasto-plastic model for granular materials with microstructural consideration. *International journal of solids and structures*, *42*(14), 4258-4277.

Chiaia, B. (2002). On the sliding instabilities at rough surfaces. *Journal of the Mechanics and Physics of Solids*, *50*(4), 895-924.

Ciavarella, M., Delfine, V., & Demelio, G. (2006). A "re-vitalized" Greenwood and Williamson model of elastic contact between fractal surfaces. *Journal of the Mechanics and Physics of Solids*, *54*(12), 2569-2591.